\documentclass[letterpaper,twocolumn,10pt]{article}
\usepackage{usenix}

\usepackage{tikz}
\usepackage{amsmath}
\usepackage{amssymb}
\usepackage{graphicx}
\usepackage{subcaption}
\usepackage{booktabs}
\usepackage{siunitx}
\usepackage{multirow}
\usepackage{longtable}
\usepackage{tabularx}
\usepackage{array}
\usepackage{url}
\usepackage{tikz}
\usetikzlibrary{arrows,positioning}
\usepackage{algorithm}
\usepackage{algorithmic}

\usepackage{filecontents}
\usepackage{titlesec}
\usepackage{hyperref}
\usepackage{xurl}
\usepackage{titlesec}

\titlespacing*{\section}{0pt}{1.2ex plus 0.4ex minus 0.2ex}{0.6ex}
\titlespacing*{\subsection}{0pt}{1.0ex plus 0.3ex minus 0.2ex}{0.4ex}
\titlespacing*{\subsubsection}{0pt}{0.8ex plus 0.2ex minus 0.1ex}{0.3ex}

\begin{document}

\date{}

\title{\Large \bf Beyond the Hype: Evaluating LLM Integration and Practical Limitations in Security Operation Centers}

\author{
Elnaz Rabieinejad\textsuperscript{1},
Ali Dehghantanha\textsuperscript{1},
Fattane Zarrinkalam\textsuperscript{2},
Sarina Dastgerdy\textsuperscript{1}\\[3pt]
\textsuperscript{1}Cyber Science Lab, Canada Cyber Foundry,
University of Guelph, Guelph, ON, Canada\\
\textsuperscript{2}College of Engineering,
University of Guelph, Guelph, ON, Canada\\
\texttt{\{erabiein,adehghan, sdastge\}@uoguelph.ca},
\texttt{fzarrink@uoguelph.ca}
}

\maketitle

\begin{abstract}
\textcolor{black}{Large Language Models (LLMs) are increasingly being explored within Security Operation Centers (SOCs) to support text-heavy analytical work such as alert contextualization, incident summarization, and drafting investigative artifacts. Despite this interest, practitioners describe critical operational concerns, most notably hallucinations (plausible but incorrect outputs), opaque reasoning, and the verification effort required to safely use model-generated content in security workflows. In this paper, we present findings from semi-structured interviews with 20 SOC practitioners spanning frontline analysts, SOC managers, and tool developers. Participants report perceived time savings for low-stakes tasks that are quickly verifiable (e.g., summarizing logs or drafting initial investigative leads), but they consistently frame LLM outputs as preliminary drafts and suggestions rather than decision-grade conclusions. Participants also describe limited trust in LLMs for high-stakes security decisions due to unreliable outputs and unclear model reasoning, and they report relying primarily on ad-hoc verification norms and continuous human oversight rather than standardized mitigation procedures. Based on these interview-grounded accounts, we introduce a maturity rubric to characterize readiness for LLM integration and outline a research agenda emphasizing auditability and transparent explanation mechanisms to support safer adoption in SOC workflows.}

\end{abstract}

\section{Introduction}
\textcolor{black}{Security Operations Centers (SOCs) face sustained overload from the growing volume and complexity of alerts, logs, and incident artifacts. Traditional Machine Learning (ML) techniques can support narrow classification and anomaly detection tasks such as intrusion detection, malware classification, and threat correlation~\cite{mink2023everybody}. However, these approaches often stop at labeling events as malicious or benign and do not consistently provide the contextual, human-readable explanations needed for fast, defensible analyst decisions under time pressure.}

\textcolor{black}{Large Language Models (LLMs) have recently been positioned as a promising response, with the ability to summarize heterogeneous incident data, generate human-readable explanations, and produce draft responses in accessible terms~\cite{Zhang2024LLMsCybersecSurvey}. This capability goes beyond detection: LLMs can translate technical evidence into narratives and candidate next steps that analysts can review, refine, and communicate. Given analyst fatigue and cybersecurity skill shortages, this automation potential is attractive to many organizations. A recent Gartner survey reflects growing industry interest in Generative Artificial Intelligence (GenAI), with approximately 55\% of companies experimenting with such technologies; however, only about 10\% report production deployment, suggesting persistent concerns about operational reliability and trustworthiness~\cite{Gartner}.}

Most prior work on LLMs in cybersecurity focuses on capability evaluations, benchmark-style studies, or broad surveys of opportunities and risks. By contrast, less is known about how LLMs are used in real SOC workflows, which workflow-grounded failure modes arise when outputs are treated as investigative leads or draft artifacts, and what organizational readiness is needed to verify outputs and mitigate hallucination risk in daily practice. In operational SOC settings, failures may be difficult to detect quickly: models can produce plausible but incorrect statements, invalid logic, or hard-to-verify explanations. Even infrequent errors can erode trust and create costly rework because practitioners must validate outputs before acting on them. Deployment is further constrained by governance requirements such as auditability, change control, and approval processes, as well as by limited access to specialized datasets, rapidly aging threat data, and computational demands~\cite{hassanin2024comprehensive}.

\textcolor{black}{In this work, we investigate how SOC practitioners perceive the utility, limitations, and mitigation strategies associated with LLM-based security tools, and we characterize when perceived benefits are bounded by failure modes and verification burden. Our study is guided by three research questions:
\begin{itemize}
    \item \textbf{RQ1:} Where and how are LLMs currently integrated into SOC workflows?
    \item \textbf{RQ2:} What are the perceived benefits and practical challenges associated with using LLMs in security operations?
    \item \textbf{RQ3:} To what extent are SOCs organizationally prepared to identify, understand, and mitigate LLM hallucination risks?
\end{itemize}}
To answer these questions, we conducted semi-structured interviews with 20 security practitioners across SOC ecosystems, including front-line analysts, SOC managers, tool developers, and researchers exploring LLM integrations. Participants described use cases, perceived benefits, failure modes, and the verification and mitigation practices they rely on in daily work.
\textcolor{black}{This scoping decision was explicit rather than post-hoc. Among the failure modes practitioners encounter, hallucination is uniquely dangerous in SOC workflows because, unlike adversarial manipulation or syntax errors, hallucinated facts and misleading investigative direction can appear plausible enough to pass initial scrutiny and silently propagate into triage checklists, detection drafts, and incident reports. We therefore scoped RQ3 around hallucination because assessing readiness to manage this risk required practitioner-grounded interview evidence that capability benchmarks alone cannot provide.}

\textbf{Scope/Non-goals.}
\textcolor{black}{We study \emph{non-adversarial operational reliability failures} when SOC practitioners use LLMs as assistants in routine workflows. Here, a \emph{failure mode} is any plausible but wrong, incomplete, misleading, or hard-to-verify output that increases workflow risk through rework, mis-triage, delay, or incorrect reporting. Our taxonomy is grounded in participant-reported day-to-day use, including hallucinated facts, invalid logic, misleading direction, interpretation errors, and verification burden, under the assumption that analysts remain accountable decision-makers. We do \emph{not} evaluate adversarial attacks or attack-surface effectiveness; these are discussed only as context because they were not directly measured in our interviews.}
\textbf{Contributions.}
\textcolor{black}{This paper makes four contributions grounded in SOC practice:
\begin{itemize}\itemsep0.2em
\item a role-stratified empirical mapping of where LLMs enter SOC workflows and which tasks practitioners delegate versus retain (RQ1);
\item a transcript-derived taxonomy of LLM output failure modes in SOC settings (FM1--FM7) with role coverage and salience (RQ2);
\item a hallucination-mitigation maturity rubric with observable criteria and a participant-level distribution (RQ3);
\item and a decision-grade, verification-first synthesis matrix connecting SOC task types, failure modes, verification burden, maturity prerequisites, and safe integration patterns.
\end{itemize}}

\textcolor{black}{The remainder of this paper is organized as follows. In Section~\ref{s2}, we review relevant literature. Section~\ref{s3} details our research methodology. We then present findings addressing our research questions in Sections~\ref{s4},~\ref{s5}, and~\ref{R3}, respectively. Section~\ref{s6} discusses implications, limitations, and actionable recommendations for future research and practice. Finally, Section \ref{s8} concludes the paper.}

\section{Related Work} \label{s2}

Recent work frames LLMs in cybersecurity as a double-edged capability, offering defensive benefits while introducing new attack risks. Focusing on operational SOC environments, we review two strands of literature: defensive SOC-oriented uses and offensive/AI-native threats, then highlight the evaluation and operational gaps motivating RQ1–RQ3.
\subsection{Defensive Uses of LLMs in Cybersecurity}
A growing body of research explores how LLMs can support security analysis tasks that are language-heavy and context-dependent, including security investigation assistance, narrative summarization, and explanation. For example, HuntGPT integrates anomaly detection and explainable AI with an LLM to support threat hunting and analyst-facing reasoning over signals \cite{Ali2023HuntGPT}. Survey work similarly highlights how LLMs are being explored across defensive tasks such as analysis, triage, and knowledge extraction \cite{Zhang2024LLMsCybersecSurvey, Zhou2024security}.

More recently, SOC-specific literature has begun to organize these capabilities around concrete SOC workflows. Habibzadeh et al. provide a comprehensive SOC-focused survey, structuring LLM use cases across core functions such as alert triage, incident response support, log analysis, and threat intelligence, while emphasizing practical constraints including data sensitivity, evaluation realism, and reliability \cite{Habibzadeh2025SOCSurvey}. Complementing surveys, early system-oriented studies examine how LLMs can be embedded into SOC toolchains. Singh et al. evaluate multi-model LLM integration with SIEM workflows for Tier-1 triage decisions, illustrating potential efficiency gains while underscoring the need for careful validation when LLM outputs affect operational prioritization \cite{Singh2024SIEMMultiModel}. Tseng et al. target a persistent SOC bottleneck, the repetitive analysis of natural-language CTI reports, and propose an LLM-based agent to automate key CTI analysis steps, aiming to reduce analyst overhead in intelligence-driven workflows \cite{Tseng2024CTIWorkflows}. Together, these works motivate our focus on where and how LLMs are integrated in real SOC practice (RQ1) and what benefits and frictions practitioners perceive (RQ2).

A second line of work focuses on developer- and code-centric security workflows. Empirical evaluations show that code assistants can introduce security risks in practice: users may accept insecure suggestions, misunderstand generated code, or over-trust outputs under time pressure \cite{Sandoval2023LostAtC, Pearce2021CopilotSecurity}. Together, these studies suggest that LLMs can be valuable as assistive tooling, but benefits are tightly coupled to verification practices, user expertise, and workflow design.

\subsection{Offensive Uses and AI-native Attack Classes}
\textcolor{black}{Prior work shows that LLM-integrated systems can be vulnerable to adversarial manipulation (e.g., malicious prompting and knowledge-source manipulation) and that LLMs may lower the cost of generating persuasive malicious content. We cite this literature to situate the broader risk landscape, not as phenomena we empirically test. For instance, prompt injection has been formalized and benchmarked as a control-plane manipulation in LLM applications \cite{Liu2024PromptInjectionBenchmark}. RAG increases reliance on external corpora \cite{Lewis2020RAG}, enabling knowledge-corruption attacks such as PoisonedRAG that can steer generations by poisoning parts of the retrieval corpus under certain conditions \cite{Zou2025PoisonedRAG}. Other work highlights LLM-enabled offensive content generation at scale (e.g., phishing) \cite{SahaRoy2024Phishbots}, jailbreak-style robustness failures across models \cite{Schwinn2023Attacks}, and risks in agentic tool-using configurations in constrained settings \cite{Fang2024OneDayVulns}. In contrast, our interview-based evidence focuses on \emph{non-adversarial operational reliability failures} in SOC workflows, incorrect or weakly grounded outputs, prompt-to-prompt variability, and the verification burden required to safely use LLM-generated drafts and preliminary leads, treating adversarial risks as background motivation rather than a measured outcome or contribution.}

\subsection{Positioning of Our Study}
\textcolor{black}{Existing work has examined where LLMs may support defenders, including SOC surveys and early SIEM/CTI automation prototypes \cite{Habibzadeh2025SOCSurvey,Singh2024SIEMMultiModel,Tseng2024CTIWorkflows}, as well as the risks these systems may introduce. However, this literature largely emphasizes technical solutions and controlled benchmarks, while giving less attention to practitioner experience, perceptions, and organizational preparedness. Prior qualitative SOC studies also describe workflows and adoption barriers, but often stop short of producing LLM-specific, workflow-grounded guidance.}

\textcolor{black}{Our study extends this literature by contributing three practitioner-derived artifacts for deployment decisions: (i) a \emph{failure-mode taxonomy}, (ii) a \emph{hallucination-mitigation maturity rubric}, and (iii) a \emph{SOC Integration Constraint Matrix} linking task types to failure modes, verification burden, maturity prerequisites, and safer integration patterns.}

\textcolor{black}{Although concerns such as hallucinations, prompt sensitivity, verification burden, and the need for human oversight are consistent with prior literature, our contribution is to show how they appear in SOC workflows specifically, where outputs are produced under time pressure, must be traceable to evidence such as logs or threat-intelligence sources, and are constrained by governance requirements. In this setting, the question is not only whether LLM outputs can be wrong, but which tasks can remain safely draft-oriented, which failures are hardest to detect quickly, and what controls are needed before broader deployment.}

\textcolor{black}{Our empirical claims are limited to non-adversarial operational reliability failures in SOC workflows; adversarial manipulation is discussed only as context. The goal is to support responsible deployment by helping teams manage hallucination-related risks in line with security requirements.}

\section{Methodology}\label{s3}

To investigate our research questions, we conducted semi-structured interviews with 20 cybersecurity professionals engaged in SOC activities. Our goal was to capture diverse perspectives on the practical use, perceived benefits, challenges, and organizational preparedness regarding the integration of LLMs in security operations. Participants were specifically selected for their direct experience in using, managing, or researching security tools within operational SOC environments.

\subsection{Semi-Structured Interviews}

Each participant completed a 30-minute semi-structured online interview conducted in English. Interviews consisted of three thematic sections, directly aligned with our research questions. Initially, participants described the integration and use of LLMs within their SOC workflows (RQ1). Subsequently, they discussed the comparative benefits and limitations of LLM-based tools versus alternative methods (RQ2). Finally, we examined participants’ perceptions of organizational readiness and strategies employed to address hallucinations in LLM outputs, specifically highlighting trust impacts and mitigation approaches (RQ3).

To maintain consistency, we developed and adhered to a standardized interview protocol, covering introductions, consent processes, clear and non-leading questions, and post-interview compensation procedures.\textcolor{black}{Before data collection, the interview guide was independently reviewed by a cybersecurity professional external to the author team and with SOC-relevant experience, in order to assess question clarity, terminology, and domain appropriateness; the guide was revised accordingly.} \textcolor{black}{In addition, we conducted a single pilot interview using the interview guide before formal data collection, and used it to refine question wording, flow, and timing.}

\subsection{Participant Recruitment}

\textcolor{black}{Participants were recruited from June to December 2024 through professional networks and platforms, notably LinkedIn, and via personal referrals. Eligible candidates had at least one year of practical SOC experience, were 18 years or older, and actively engaged with cybersecurity operations. We did not require explicit expertise in AI or LLMs, allowing us to include varied perspectives reflective of the broader SOC landscape. This recruitment strategy, similar to prior qualitative studies in cybersecurity~\cite{mink2023everybody,dietrich2018investigating, alahmadi202299} which had below 20 participants, do not claim our findings are generalizable given the sample size. Instead, we use the data to highlight salient emerging themes and concepts. All interviews were audio-recorded with participant consent and transcribed verbatim by a GDPR-compliant professional service. Each participant received a \$50 CAD electronic gift card upon completion. }
Participant demographics are summarized in Table~\ref{t1} in Appendix \ref{b1}, highlighting a wide range of roles, experiences, and organizational contexts represented in our sample.

\subsection{Data Analysis}

Interview recordings were transcribed verbatim by a General
Data Protection Regulation (GDPR)-compliant professional
service and analysed using codebook thematic analysis fol-
lowing Braun and Clarke’s phases (familiarisation, generat-
ing codes, searching/reviewing themes, defining/naming, re-
porting)~\cite{braun2006using}. Two researchers
independently coded an initial subset (5 interviews) to develop a shared
codebook, compared interpretations, and reconciled discrepancies.
Inter-rater reliability (IRR) was then estimated on held-out batches every
few interviews using Cohen’s $\kappa$, with interpretation following
Landis and Koch~\cite{landis1977measurement}. Coding proceeded iteratively
with periodic codebook updates; when agreement fell below our threshold
($\kappa < 0.80$), we resolved differences by discussion, refined code
definitions, and retro-fit earlier transcripts as needed. After five coding
rounds, agreement stabilised at $\kappa > 0.80$ across thematic categories.
Cohen’s $\kappa$ computed at the subcode level on held-out transcripts
improved across rounds from $\kappa = 0.81$ to $0.88$ [Round~1: 0.81;
Round~2: 0.83; Round~3: 0.85; Round~4: 0.86; Round~5: 0.88].
The most frequent disagreements were boundary cases between \emph{data
leakage/model elicitation} vs.\ \emph{prompt/model misuse}. Consistent with
our recruitment rationale, we monitored both code and meaning saturation
during analysis and observed no new substantive codes after Interview~15,
continuing to $N{=}20$ to strengthen meaning saturation. We define saturation
as no new sub-themes or dimensions emerging in the last five transcripts,
with no changes to code definitions~\cite{hennink2017saturation}. The final codebook,
exemplar quotes, and IRR statistics appear in Appendix \ref{a1} (Table~\ref{t2}
and Table~\ref{t3}).

\textcolor{black}{According to Table \ref{tab:code_saturation_batch} in Appendix \ref{c1}, to demonstrate the internal robustness of our thematic discovery, we conducted a saturation audit by tracking the first appearance of each code within our participant batches. We defined saturation as the point where a batch of five interviews yielded no new top-level primary codes and fewer than 5\% new subcodes. }

\begin{table}[t]
\centering
\small
\color{black}
\caption{Code saturation by participant batch.}
\label{tab:code_saturation_batch}
\setlength{\tabcolsep}{3pt}
\renewcommand{\arraystretch}{1.15}
\begin{tabular}{l r r r r}
\hline
\textbf{Interview Batch} &
\textbf{New Primary} &
\textbf{New} &
\textbf{Cumulative} &
\textbf{\% Final} \\
& \textbf{Codes} & \textbf{Subcodes} & \textbf{Codes} & \textbf{Codebook} \\
\hline
P01--P05  & 14 & 108 & 122 & 82\%  \\
P06--P10  & 0  & 18  & 140 & 94\%  \\
P11--P15  & 0  & 9   & 149 & 100\% \\
P16--P20  & 0  & 0   & 149 & 100\% \\
\hline
\end{tabular}
\end{table}

\textcolor{black}{All 14 primary thematic categories (e.g., Challenges \& Limitations, Adaptability, Comparison to Rule-Based) were established within the first five interviews. Between Batch 2 and Batch 3, the emergence of new subcodes dropped from 18 to 9, primarily consisting of niche job roles or specific tool names rather than new conceptual challenges.In the final batch (P16–P20), no new codes or dimensions were added to the codebook, confirming that our sample of $N=20$ was sufficient to capture the breadth of practitioner experiences in this regional SOC ecosystem.} \textcolor{black}{We classify participants as LLM users if they reported any workplace LLM use (including occasional or low-intensity use). Participants who reported no workplace LLM use are labeled Not applicable because the mitigation maturity rubric is not applicable to them. Statements from participants without approved workplace LLM use are treated as perceptions or indirect exposure and are not used to infer operational deployment.}




\textcolor{black}{For all thematic findings, the individual participant ($N=20$) serves as the primary unit of analysis. While participants represent 13 distinct organizations, their reported experiences reflect individual roles and specific sub-teams. Consequently, maturity levels and usage patterns are classified at the participant level to capture the heterogeneity of practice within larger organizations. Where a metric is reported at the organizational level (e.g., demographics in Section \ref{s41}), it is explicitly noted.}\\
\textcolor{black}{\textbf{Researcher positionality}. The research team has experience in cybersecurity research and SOC-adjacent problem settings, which informed the framing of the interview protocol and interpretation of practitioner terminology. To reduce interpretive bias, we used independent coding, iterative codebook refinement, and consensus-based reconciliation throughout the analysis.}

\subsection{Limitations}
Our study has several limitations typical of qualitative work. First, while semi-structured interviews enabled depth, response detail varied because follow-up probing was not perfectly uniform across participants; core questions were covered consistently, but some themes have richer narrative evidence than others. \textcolor{black}{As with most interview-based qualitative studies, our findings reflect participant accounts rather than direct observation of workflows, creating the possibility of self-reporting and social desirability effects.}

Second, the sample is geographically concentrated (19/20 in Ontario, Canada; 1 in California, USA). Although participants represented organizations of varied sizes (Table~\ref{t1}), we do not claim global representativeness, nor did we systematically control for SOC maturity beyond self-described roles and contexts. The results should therefore be read as an in-depth regional snapshot of early LLM use and as practice-informed hypotheses, not sector-wide benchmarks. \textcolor{black}{Gender was collected as a demographic variable (16 male, 4 female) but was omitted from Table~\ref{t1} to preserve participant anonymity given the small sample size; the gender imbalance toward male participants may limit the breadth of perspectives captured.}

Third, confidentiality and NDAs meant we did not require disclosure of LLM vendors, model versions, deployment mode (hosted vs.\ on-prem), or fine-tuning, so we cannot separate model/deployment-specific effects from broader adoption patterns; our findings capture use patterns and practitioner perceptions, not comparative performance across models. Fourth, the findings are time-stamped to mid--late 2024, before more integrated agentic workflows and widespread grounding became common; while tools may evolve quickly, SOC governance and verification typically change more slowly, so we frame our themes as a baseline for future longitudinal study. 
\textcolor{black}{In addition, our coding framework allowed failure modes to co-occur through multi-label coding, but co-occurrence reporting was not a pre-specified objective of this paper. We therefore do not quantify co-occurrence frequencies here, and leave systematic analysis of the most common co-occurring pairs for future work.}
Finally, we do not measure operational KPIs (e.g., incident response time, FP/FN reduction, detection lift); claims are limited to interview-grounded descriptions of practice and perceived utility, usually conditional on verification.

\subsection{Analytic Triangulation and Robustness Checks}
\textcolor{black}{We employed two internal robustness checks to ensure our findings were not biased by our initial analytical lenses:}

\textcolor{black}{\textbf{Negative-Case Analysis:}} 
\textcolor{black}{While most participants reported a heavy verification burden, outliers like P12 and P17 highlighted a critical boundary condition: LLM utility is highly task-dependent. In code-centric generative tasks, such as drafting filtering logic or Python scripts, practitioners experienced minimal friction because outputs are rapidly falsifiable through technical tools like execution tests and syntax checks. Conversely, skepticism remained high for investigative or narrative-driven tasks—such as incident reasoning or attribution—where errors often appear plausible and are difficult to verify quickly, necessitating much heavier human judgment.}

\textcolor{black}{\textbf{Sensitivity Analysis:} }
\textcolor{black}{To confirm that our conclusions are not an artifact of presentation structure, we re-aggregated the coded segments using two alternative organizational lenses: (i) a task/workflow lens (Section \ref{s4}) and (ii) an evaluation-criteria lens (Section \ref{s5}). We treat the task/workflow lens as primary, and use evaluation criteria only as a supporting organization for how participants justified trust and perceived utility. Across both lenses, the core conclusion remained unchanged: perceived LLM utility is highest in text-centric, draft-oriented tasks, while adoption is bottlenecked by verification burden and limited trust in decision-grade outputs.}

\section{Results: Tool Usage (RQ1)}\label{s4}

In this section, we first provide an overview of participants’ professional backgrounds and then present findings on how LLM-based security tools are integrated into SOC workflows, using LLM output failure modes (i.e., incorrect or untrusted outputs) as the central lens for understanding readiness and trust in this integration. \textcolor{black}{ Our results reflect two kinds of interview-grounded evidence: (i) reported practices (how participants say they currently use LLMs in SOC workflows, including what artifacts they generate and how they validate them), and (ii) perceptions (participants’ reported benefits, costs, and trust judgments). We do not evaluate decision-grade operational outcomes (e.g., detection lift, false positive reduction, or incident response time reduction). Accordingly, terms such as “effectiveness” and “efficiency” in this section refer to participant-reported perceived utility and workflow support, typically conditional on verification.}

\subsection{Participant Background and Demographics}\label{s41}

Table~\ref{t1} summarizes demographics and professional profiles for 20 participants spanning diverse cybersecurity roles (front-line analysts, SOC managers, threat-response engineers, data scientists, and senior leadership) with 1.5–25 years of experience. We classified participants as \emph{SOC professionals} if their primary work involved day-to-day SOC operations (e.g., triage, incident response, detection engineering, or SOC management), and as \emph{security researchers} if they primarily worked in research/R\&D but had direct SOC workflow involvement within the past 12 months; purely academic researchers without operational SOC engagement were excluded. Participants’ expertise ranged from defensive-only to combined offensive/defensive practice, with varied methodological preferences (rule-based, AI-driven/ML/LLM, or hybrid) and wide-ranging AI experience from conceptual familiarity to hands-on model development. The sample was geographically concentrated (19/20 Ontario, Canada; 1 California, USA) and drawn from 13 organizations spanning in-house enterprise SOCs, public-sector SOCs, MSSPs/providers, and security product vendors across all scale bands—small ($50$–$249$), medium ($250$–$999$), large ($1{,}000$–$9{,}999$), and very large ($\ge 10{,}000$); to preserve anonymity, we report scale bands rather than exact headcounts, and no single organization dominated the sample.

\subsection{Common Use-Cases and Contextual Dependencies}

\textcolor{black}{Across participant categories, four use-cases appeared consistently: log summarization, initial incident investigation support, documentation drafting, and detection rule refinement. Rather than treating these as task types alone, we characterize each as a workflow segment with an identifiable entry point, decision gate, and transition condition, based on participant accounts.}

\textcolor{black}{\textbf{Log summarization.} Entry was volume-driven: analysts invoked LLMs when a log artifact exceeded what could be parsed manually at triage speed (e.g., P09, P10). After prompting for a summary, they applied a rapid plausibility check, comparing the output against a spot-check of the raw logs, before deciding to escalate, close, or continue. The falsification condition is fast and concrete (``is this entity in the raw log?''), which explains why participants reported low verification burden for this task.}

\textcolor{black}{\textbf{Triage guidance.} A second entry point was unfamiliarity with an alert type. Analysts (e.g., P13) described prompting for first-pass investigation steps, then filtering the output against internal playbooks, retaining steps supported by available telemetry and discarding those that contradicted known-safe behavior or lacked log evidence. The LLM output served as a brainstorm scaffold, not an executable runbook; each suggested step required independent justification before execution.}

\textcolor{black}{\textbf{Detection rule drafting.} Engineers (e.g., P04, P12) invoked LLMs after a detection gap was already identified, providing a natural-language description or a suspicious log excerpt and receiving a candidate query or rule fragment. The immediate decision gate was syntactic: linting or sandbox execution to confirm operator and field validity. Rules that passed moved to backtesting against historical traffic before peer review and production promotion. Failures, including the FM2 case where an LLM invented nonexistent operators (P04), triggered manual correction or prompt revision.}

\textcolor{black}{\textbf{CTI and report synthesis.} Researchers (e.g., P07, P19) invoked LLMs after aggregating intelligence from multiple feeds, prompting for TTP extraction or a structured summary. The decision gate was cross-referencing: extracted indicators were checked against internal threat intelligence platforms or MITRE ATT\&CK before operational use. P07 described constraining prompts to provided source text to reduce inference-driven fabrication.}

\textcolor{black}{A consistent structural pattern emerged across all four cases: the LLM is invoked at a specific handoff point in an existing workflow, and its output passes through a human-operated decision gate before any downstream action. The feasibility of that gate, and therefore the practical verification burden, depends on whether the output type admits a fast, concrete falsification condition. This structure motivates the verification burden and maturity prerequisite mappings in the SOC Integration Constraint Matrix (Appendix~\ref{app:synthesis_matrix}).}

\section{Results: Perception (RQ2)} \label{s5}

We asked participants to describe their experiences and perceptions regarding the integration of LLM-based tools into SOC workflows, specifically focusing on the perceived benefits and challenges. From the thematic analysis of our interviews, five distinct evaluation criteria emerged consistently: \emph{Effectiveness}, \emph{Usability}, \emph{Efficiency}, \emph{Adaptability}, and \emph{Trust}. Below, we detail the participants' insights with respect to these criteria, emphasizing particularly prominent themes.

\subsection{Effectiveness}
\textcolor{black}{Effectiveness refers to an LLM's capacity to generate accurate and contextually relevant outputs that support operational security tasks~\cite{zhou2021evaluating}. Participants consistently described perceived workflow benefits in some SOC tasks, particularly where outputs were draft-oriented and quickly verifiable.}

\subsubsection{LLMs Excel in Text-Centric Tasks}
\textcolor{black}{A dominant perception among participants (13 of 20) was that LLMs can provide useful workflow support in text-intensive SOC tasks by reducing the time and manual effort associated with summarizing logs and reports, drafting standardized detection content, and producing documentation. Participant P09, for example, described rewriting detection rules with LLM assistance as “much faster” than manual drafting, while still treating the output as a draft requiring review. Participants further suggested that automating routine language-related tasks could allow analysts to redirect attention toward more complex activities requiring human expertise. We interpret these accounts as participant-reported perceived utility and workflow support, rather than as direct evidence of measured performance improvement or operational productivity gains.}

\subsubsection{Emerging Use in Detection Engineering Support}
A subset of participants described using LLMs for detection engineering support, where models help analysts draft and refine artifacts that may later feed existing detection workflows. Reported uses included extracting salient fields from log excerpts, drafting triage checklists, and generating candidate query/rule fragments or investigative steps. A smaller subset discussed experimental pilots in which LLMs proposed candidate detection ideas from log snippets, but these outputs were consistently framed as suggestions that require rigorous human validation and conventional backtesting/review before any operational adoption. Overall, participants viewed LLMs as accelerating early-stage drafting and hypothesis generation, not replacing detection pipelines or analyst judgment.

Three participants also mentioned experimental deployments where LLMs directly suggested detections from log excerpts or suspicious data points; again, these suggestions were universally described as needing strict human verification prior to acceptance. Together, the findings indicate a cautious but growing integration of LLMs into SOC detection work, balancing the speed benefits of automation with ongoing accuracy concerns and the continued necessity of analyst oversight, with an envisioned model where LLM tools augment rather than replace human analysts.

\subsection{Usability}
Usability refers to the ease with which SOC analysts can effectively interact with and configure LLM-based tools within their existing workflows~\cite{zytek2021sibyl}. While participants generally appreciated the intuitive nature of chat-like interfaces compared to traditional, specialized ML toolchains, they also highlighted critical usability challenges, particularly related to prompt engineering and transparency of outputs.

\subsubsection{Accessible Interfaces Accelerate Onboarding}
\textcolor{black}{Participants widely recognized the accessibility and intuitive
use of conversational interfaces, particularly highlighting widely
available platforms such as ChatGPT. }Seven participants explicitly praised user-friendly features such as conversational histories, natural language prompts, and proactive disclaimers about potential inaccuracies. Another four participants reported primarily interacting with LLM outputs via integrated, vendor-based security platforms (e.g., Splunk, Microsoft Defender), appreciating the seamless embedding of LLM functionalities within familiar security dashboards.

A frequently cited advantage (n=3) was the minimal technical setup required to initiate interaction with LLM-based tools. Participants described significant ease in formulating initial queries, detection rules, or summaries without needing custom code. For example, P12 emphasized this usability benefit by stating: \textit{``I don't need to code intricate filters; I just describe what I want, and the tool drafts it.''} The approachable interface reduces barriers to adoption, enabling rapid onboarding and reducing barriers to drafting and exploratory tasks (e.g., writing initial summaries or first-pass rule/query drafts), especially for analysts who can rapidly validate outputs.

\subsubsection{Challenges with Prompt Engineering}
Despite initial ease of use, eight participants highlighted prompt engineering as a key usability challenge. They noted that vague or inadequately specified prompts frequently resulted in overly general or irrelevant outputs, requiring repeated adjustments or clarifications. Participant P07 emphasized the effectiveness of incorporating real-world examples or log excerpts directly into prompts, stating that this practice significantly improved output accuracy.

This finding indicates an inherent tension between the superficial simplicity of LLM interfaces and the nuanced skills required to craft effective prompts. While the chat-based interaction model appears straightforward, achieving consistent, relevant results heavily depends on analyst domain knowledge and precise query formulation. Thus, successful integration and sustained usability of LLM-based tools depend substantially on analysts’ expertise and their familiarity with effective prompting techniques.

\subsubsection{Nontransparent Reasoning Hinders Adoption}
Seven participants raised significant usability concerns related to the opacity of LLM reasoning processes. Unlike traditional rule-based detection methods, where triggered rules explicitly clarify why certain outputs are generated, LLM outputs are often presented without clear justification or traceable reasoning. Participant P05 described this concern vividly: \textit{``With rule-based systems, at least we know which rule triggered. The LLM just gives an answer, and I'm not always sure why.''} 

Such opacity frequently compels analysts to conduct supplementary manual verification—examining logs, consulting historical records, or otherwise attempting to infer the rationale behind generated outputs. This lack of transparency not only hampers trust in LLM-generated results but can also slow incident response times, limiting overall operational efficiency.

In summary, while the initial usability of LLM-based tools is positively perceived, practical adoption is significantly influenced by challenges around effective prompt engineering and output transparency. Addressing these issues through improved user training, clearer explanatory mechanisms, or enhanced interface designs could substantially enhance the usability and operational utility of these tools in SOC environments.

\subsection{Efficiency}
Efficiency refers to how LLM-based solutions help analysts complete security tasks with minimal time, data, and computational overhead~\cite{zhou2021evaluating}. Participants consistently highlighted reported perceived efficiency gains—particularly in automating repetitive tasks—but also emphasized the importance of balancing these gains against verification overhead.

\subsubsection{Time-Saving and Workload Reduction for Repetitive Tasks}
A majority (n=13) of participants emphasized the substantial time-savings enabled by LLM assistance in routine, language-intensive SOC tasks such as log summarization, incident documentation, and security briefings. For instance, P16 noted that email security summaries previously taking ``an hour or two to compile'' were now generated ``within minutes.'' Three participants explicitly mentioned workload reductions, indicating that LLMs effectively reduced the burden of mundane tasks, enabling them to focus more on complex investigations. 

Despite these advantages, participants acknowledged that effective prompt engineering was still necessary to ensure coherence and relevance in LLM-generated outputs. These accounts suggest participants perceived workload relief in specific tasks, especially where outputs were quickly verifiable.

\subsubsection{Automation Potential Beyond Text Summarization}
Beyond text-based tasks, five participants recognized the potential of LLMs to facilitate semi-automated scripting and investigation workflows. Participant P17, for example, described using LLM-generated Python code snippets for rapid log analysis, reducing manual coding effort in their workflow, conditional on validation. Similarly, participant P13 mentioned that LLMs effectively drafted ``first-pass triage steps,'' enabling the team to concentrate on deeper forensic analysis. Nevertheless, all participants using such semi-automated workflows stressed the essential role of human review due to the persistent risk of inaccuracies and logical errors in automated outputs.

This underscores an emerging hybrid model, where LLMs accelerate initial task stages but require analyst-led validation, balancing efficiency gains with accuracy requirements.

\subsubsection{Verification Overhead May Erode Efficiency Gains}
Despite general satisfaction with accelerated content generation, several participants expressed concerns about the associated verification burden. Participant P04 stated explicitly, ``We'll save an hour drafting rules but spend half an hour verifying every line.'' This overhead becomes especially prominent in critical situations—such as near-real-time incident response or high-visibility reporting—where thorough verification is essential. Many participants described cross-referencing LLM outputs with internal procedures or historically validated rules to mitigate risks.

This verification necessity highlights a fundamental aspect of LLM deployment in security: although the speed advantages are clear, ongoing human oversight remains crucial. Thus, participants generally perceived efficiency benefits as contingent upon their capacity to quickly identify and correct occasional inaccuracies, particularly in low-risk contexts.

\subsection{Adaptability}
Adaptability describes an LLM-based tool's capability to seamlessly integrate into existing organizational processes and respond effectively to varying operational contexts~\cite{zhou2021evaluating}. While participants recognized the general adaptability of LLMs, they highlighted notable limitations when encountering novel or highly specialized security scenarios.

\subsubsection{Flexibility Across Data Sources and Tasks}
Several participants (n=6) positively assessed the adaptability of LLMs across diverse SOC tasks and data sources, including asset risk scoring, summarizing alerts, and synthesizing vulnerability metrics. Participant P17, for example, successfully integrated LLM outputs with internal classification data, noting that the model ``shifts easily between different environments if given the right inputs.'' However, two participants clarified that effective adaptation often required significant domain-specific input and careful data curation from knowledgeable analysts.

This suggests that LLM adaptability relies heavily on ongoing analyst input, highlighting the importance of domain expertise and proactive context provisioning in sustaining successful deployment.

\subsubsection{Limits with Novel or Specialized Threats}
Despite overall versatility, four participants expressed concerns regarding LLM performance deterioration when dealing with novel or specialized threats, such as zero-day vulnerabilities lacking historical references. Participant P19 described that \textit{``when it's a fresh zero-day or we have minimal intel, the LLM is forced to guess, and sometimes that guess is off-base.'' }Consequently, in highly specialized scenarios, LLM predictions often become error-prone.

Participants indicated two primary strategies to mitigate this issue: quickly supplementing models with specialized threat data or reverting to manual investigations. However, three participants raised concerns about resource constraints, noting that continuously fine-tuning models to emerging threats is expensive, time-consuming, and frequently limited by internal expertise.

\subsubsection{Organizational Barriers and Data Policies}
Beyond technical considerations, seven participants noted significant organizational barriers affecting adaptability. Challenges included compliance regulations, data governance policies, vendor approval processes, and leadership hesitancy. Participant P15 characterized deploying an internal LLM as a ``multi-layer labyrinth'' requiring extensive approvals from Legal, Privacy, and Compliance teams. Conversely, two participants noted smoother adoption experiences when senior management explicitly supported AI initiatives, facilitating faster piloting and experimentation.

Budget constraints further impacted adaptability: Participant P07 observed that frequent model fine-tuning was costly, causing some organizations to rely solely on general-purpose, vendor-trained models. Thus, practical adaptability hinges not only on technical flexibility but also on organizational structures, resource availability, and policy alignment.

\subsection{Trust}
Trust refers to users' confidence in the reliability and factual accuracy of LLM outputs, particularly regarding resistance to inaccuracies and fabricated information~\cite{zhou2021evaluating}. Participants consistently identified trust as a critical factor influencing LLM adoption in SOC environments.
\subsection{Taxonomy of LLM Output Failure Modes \label{sec:failure-mode-taxonomy}}
We move beyond treating hallucination as a single issue by defining a practitioner-grounded taxonomy of LLM output failure modes observed in SOC work, based entirely on our interview corpus. We focus on failures that produce incorrect, ungrounded, invalid, or hard-to-verify outputs that could be adopted in routine SOC tasks (e.g., detection drafting, interpreting suspicious activity, investigative guidance). The taxonomy reflects practitioner-described breakdowns and verification burdens, and intentionally excludes adversarial threats (e.g., prompt injection) except where participants reported first-hand incidents.

Using coded segments about trust failures and incorrect outputs, we identify seven failure modes and assign severity via a consequence-if-acted-upon rubric: Low (wasted time/minor confusion), Medium (misdirected triage/investigation and forensic drag), and High (plausible missed detections, incorrect escalations, or unsafe actions such as blocking legitimate traffic). These severities represent conceptual potential consequences, not observed incident impact in our study.

Appendix Table~\ref{tab:failure-mode-taxonomy} (Appendix \ref{c1}) reports the taxonomy with inclusion/exclusion criteria and representative evidence. For reproducibility, we use thematic segments (sentence groups describing a specific failure event) as the unit of analysis and apply multi-label coding for compound failures (e.g., FM1 + FM5). Appendix Table~\ref{tab:taxonomy-coverage} (Appendix \ref{c2}) summarizes coverage and role-based salience across participants.
Collectively, this taxonomy clarifies that what practitioners describe as hallucination in SOC settings spans several distinct failure modes, each with different operational outcomes. This structured approach enables more precise recommendations by tying mitigations (e.g., schema validation for FM2 or RAG for FM1) to specific failure classes rather than to an unquantified notion of trust.


\subsubsection{Essential Role of Human Oversight}
Nearly all participants (n=15) explicitly reinforced the necessity of ongoing human oversight. While many participants described perceived workflow benefits in specific tasks from LLM assistance, participants insisted that human judgment was indispensable for verifying outputs. Participant P08 succinctly stated:\textit{``an LLM is an assistant, never the final authority,''}  reflecting a prevailing sentiment that analysts must thoroughly vet LLM-generated detections, summaries, and recommendations.

This emphasis on human oversight aligns with an emerging best practice in cybersecurity: combining AI-driven acceleration with expert human scrutiny. Participants described a model of use where any potential workflow benefit depends on keeping analysts in the loop to validate outputs and prevent errors from propagating.

In summary, although participants acknowledged the perceived operational usefulness provided by LLMs, they remain cautious about fully entrusting AI with critical security decision-making tasks. Near-term trust appears fundamentally dependent on robust oversight and validation procedures capable of managing the inherent uncertainty in current LLM technologies.

\section{Results: Readiness (RQ3)} \label{R3}
\textcolor{black}{In this section, we examine how participants described the readiness of their SOC workplace contexts to manage hallucination-related reliability risks associated with LLM outputs in SOC workflows, drawing on the failure-mode taxonomy introduced in Section~\ref{sec:failure-mode-taxonomy} (Table~\ref{tab:failure-mode-taxonomy} in Appendix \ref{c1}). Our findings reflect participant-reported practices (unit of analysis: participants, N=20), not organization-level maturity estimates. While participants widely acknowledged these risks as a core barrier to trustworthy integration, they described a fragmented set of readiness practices, with limited formalized countermeasures and heavy reliance on human validation.}

\textbf{Non-adoption as a boundary condition.}\textcolor{black}{In our corpus, non-adoption was analytically meaningful rather than a simple absence of use. Participants who reported no approved workplace LLM use, or who described non-adoption in their teams, associated that boundary with concerns about output reliability, the absence of approved pathways for handling sensitive data, governance restrictions on how such tools could be used in operational environments, and uncertainty that any efficiency gains would justify the added verification burden. We therefore interpret non-adoption as evidence about bounded readiness conditions in early SOC adoption, rather than as a missing case or lack of relevance.}

\textbf{Operational risk model.}\textcolor{black}{To make our security framing explicit and falsifiable from interview evidence, we operationalize hallucination risk as a set of observable outcome types for an LLM output used during a SOC task. We treat an output as a failure when it exhibits one or more of the following: (i) fabricated facts/entities (e.g., fictitious incident details or nonexistent IOCs/TTPs), (ii) schema- or tool-invalid artifacts (e.g., invalid detection-rule syntax or invented operators), or (iii) misleading procedural recommendations (e.g., incorrect first-pass triage steps or investigative guidance that increases verification burden or steers analysis incorrectly).}

\textcolor{black}{A failure becomes operationally consequential (i.e., “successful” in risk terms) when it is \emph{plausible enough to be acted upon or treated as a lead}, when it \emph{causes wasted analyst time} through rework and verification, or when it \emph{propagates error} into an incorrect investigative step, draft rule, or report content. Importantly, we do not measure incident impact or business loss, and we do not evaluate attacker adaptation or adversarial robustness; instead, we report practitioner-described failure modes and the mitigation practices they use in operational settings. Table~\ref{tab:hallucination-risk-model} in Appendix \ref{c1} summarizes these outcome types, the operational consequences that make them risky in practice, and the mitigation practices reported by participants.}

\subsection{General Observations About Hallucinations}

\textcolor{black}{Participants consistently recognized hallucination as a unique operational failure mode distinct from traditional false positives generated by rule-based systems or conventional ML models.} Specifically, they noted that LLMs can fabricate extensive narratives—such as incident details, fictitious indicators of compromise, or nonexistent tactics, techniques, and procedures (TTPs)—posing a significant risk to SOC workflows. Participant P06 captured this sentiment succinctly: \textit{``Most of us know LLMs can make things up, but we’re not sure how to systematically prevent it. Our current approach is basically: ‘keep your eyes open, and hope no one junior takes it at face value.’'} Although participants universally refrained from letting LLM outputs directly influence high-stakes decisions without human oversight, the absence of automated or robust verification processes often led to manual, resource-intensive triage efforts. \textcolor{black}{These accounts map directly to the operational outcome types in Table~\ref{tab:hallucination-risk-model}, reinforcing that the risk we study is grounded in observable workflow failures rather than adversarial threat coverage.}

\subsection{Specific Tactics for Handling Hallucinations}

Participants reported employing several distinct tactics—ranging from informal manual verification to experimental automated solutions—to address hallucination risks. Below, we summarize these strategies according to their prevalence and maturity.

\subsubsection{No Formal Mitigation Method}
\textcolor{black}{Over one-third (n=8) of participants reported that, in their immediate work setting, they did not observe any formalized strategy for managing hallucinations. Instead, hallucinations were handled ad hoc, relying on analyst judgment and manual cross-checks.} Participant P14 exemplified this approach: \textit{``If we see something that doesn’t align with the logs or feels off, we just know to question it.''} Participants attributed this absence of structured methods to limited organizational resources, risk aversion toward autonomous AI, and uncertainty regarding effective mitigation practices. Consequently, hallucination identification depended largely on analyst vigilance and intuition rather than defined protocols, further increasing senior analysts' workload and verification burden.

\subsubsection{Heavy Reliance on Human Oversight}
The most common approach reported (n=10) involved extensive reliance on human oversight. Analysts frequently cross-referenced suspicious LLM-generated outputs against internal knowledge bases, log repositories, and external threat intelligence platforms. Participant P09 summarized this practice: \textit{``We don’t have an automated safety net, so we treat the LLM output as a lead, not as a conclusion.''} Analysts described themselves as human validators, responsible for manually verifying and correcting questionable outputs, a process participants acknowledged was resource-intensive and reduced the efficiency benefits initially anticipated from LLM-based automation.

\subsubsection{Prompt Refinement}
Three participants explicitly discussed prompt refinement as an effective strategy to proactively minimize hallucinations. Rather than allowing LLMs unrestricted freedom in generating narratives, these participants carefully crafted prompts to constrain responses—for instance, using specific instructions such as bullet points or restricting content solely to provided data. Participant P07 noted a marked reduction in errors when instructing the model explicitly: \textit{``I told the LLM, ‘never assume anything we haven’t provided in the logs,’ and the error rate plummeted.''} However, participants acknowledged that this approach increased analysts’ cognitive overhead, as effective prompt engineering required specialized skill sets beyond typical SOC analyst competencies.

\subsubsection{Self-Correcting Loops}
A minority of participants (n=2) reported experimental attempts at employing self-correcting loops, where the LLM reviews and assesses its outputs to identify inconsistencies or contradictions. Although these methods occasionally flagged logical errors, participants remained skeptical of relying solely on a single model for self-validation, cautioning that \textit{``it can just hallucinate a second time.''} However, they suggested that combining self-correcting loops with RAG from trusted sources might offer future potential.

\subsubsection{Confidence Factors and Cross-Model Checks}
Few participants explored advanced validation methods involving confidence scores or multiple model cross-checks (n=2). One participant integrated LLM-generated confidence metrics with SIEM metadata, while another advocated parallel querying of multiple LLMs to detect contradictory outputs. For instance, participant P17 stated: \textit{``If my local LLM says an IP is malicious but ChatGPT says it’s benign, I investigate further.''} Although such redundancy occasionally identified inaccuracies, participants acknowledged this approach as resource-intensive and not foolproof against consistent hallucinations across models.

\subsubsection{Alignment with Established Standards}
One participant explicitly attempted to validate LLM outputs by referencing recognized security frameworks, such as MITRE ATT\&CK or NIST guidelines. If the LLM-generated information lacked identifiable links to these standards, the analyst treated the output as suspect. Nevertheless, this method required ongoing manual verification, as noted by the participant: \textit{``I still had to check if the MITRE technique ID it gave me was made up.''}

\subsubsection{Structured Output Validation}
Two participants described early-stage initiatives aimed at validating LLM outputs against authoritative data schemas or internal databases. One employed scripts that automatically cross-referenced IP addresses, hosts, or file hashes with trusted internal databases. Another converted LLM responses into structured JSON for automated schema validation, preventing syntax errors or nonexistent operators from infiltrating SOC workflows. While these structured validations effectively captured basic syntax-level errors, participants admitted deeper factual inaccuracies still required human scrutiny.

\subsection{Hallucination Mitigation Maturity Rubric}
We define a Hallucination Mitigation Maturity Rubric to replace informal readiness assessments with a transcript-evidence–based classification of hallucination-mitigation practices (Table~\ref{tab:hallucination-mitigation-maturity}, Appendix~\ref{c1}). Using participants as the unit of analysis ($N{=}20$), we coded each participant’s described practices for constraining, validating, and governing LLM outputs (e.g., prompt constraints, review gates, automated checks) and assigned them to the highest maturity level they explicitly described as a routine practice (personally or as an org/team procedure). The rubric spans four levels: L0 = ad-hoc individual verification; L1 = shared-but-undocumented norms (e.g., routine manual cross-checking); L2 = at least one documented, consistently used preventive guardrail (e.g., prompt libraries/templates or evidence-only prompting); and L3 = at least one proactive, auditable control beyond manual review (e.g., automated validation scripts, schema validation, formal Legal/Compliance sign-off gates, or audit logging of verification). Participants with no active workplace LLM use (so mitigation could not be assessed) are labeled Not applicable and excluded from preparedness claims.
\textcolor{black}{Applying this rubric revealed a distribution skewed toward lower maturity levels: Level~1 was most prevalent (30\%, $n{=}6$), followed by Level~0 (20\%, $n{=}4$). Fewer participants reported practices consistent with Level~2 (15\%, $n{=}3$) or Level~3 (20\%, $n{=}4$). Three participants (15\%, $n{=}3$) reported no workplace LLM use and are therefore labeled \textit{Not applicable } (Table~\ref{tab:hallucination-mitigation-maturity}). Each level reflects the share of participants whose transcripts described practices consistent with that level; it should not be interpreted as an organization-level readiness estimate across the 13 organizations.}

\subsection{Stratified Analysis: Operational Perspectives by Role}
\label{sec:stratified-role-analysis}

\textcolor{black}{To provide actionable design implications, we conducted a stratified analysis of our findings across four key professional roles identified in our metadata: \emph{Front-line Analysts}, \emph{Technical Engineers}, \emph{Security Managers}, and \emph{Tool Developers/Researchers}. Table~\ref{tab:stratified-role-analysis} summarizes how these roles differ in their usage, risk perceptions, and mitigation priorities.}

\textcolor{black}{The study concludes that a one-size-fits-all AI strategy is ineffective for security centers, as different roles demand distinct tool functionalities. To succeed, designers must address three key areas: balancing analyst requirements for speed and sourcing against managerial needs for policy compliance and privacy; distinguishing between tasks suitable for AI—like generative report drafting where verification is easy—and high-stakes investigative reasoning where AI opacity remains a barrier; and providing role-specific features, such as side-by-side data comparison for analysts and audit trails for leadership. These specialized focus areas are essential for navigating the current challenges and the future evolution of AI-human collaboration in security operations.}

\subsection{Decision-Grade Synthesis: SOC Integration Constraint Matrix}
This subsection introduces a single actionable artifact for SOC designers: the SOC Integration Constraint Matrix (Appendix~\ref{app:synthesis_matrix}, Table~\ref{tab:synthesis_matrix}). The matrix summarizes design constraints by mapping common SOC task types to (i) dominant failure modes (FM1–FM7), (ii) verification burden (low/medium/high), (iii) minimum mitigation maturity required (L0–L3), and (iv) a safe integration pattern (e.g., chat-only drafts, template+schema validation, gated deployment). It is populated solely from coded interview evidence: two coders independently performed the mappings and resolved disagreements via the same consensus protocol as thematic coding; unclear cases are marked Insufficient Evidence rather than extrapolated. The matrix highlights a key boundary: tasks with quickly falsifiable outputs (e.g., scripts/rules) can be integrated with lower verification effort, while narrative/investigative decision-support requires higher maturity controls and stronger gating because errors are harder to detect quickly.

\section{Discussion} \label{s6}
This study examined LLM integration into SOCs, offering insights into practitioners' perceptions and usage. While LLMs show potential for automating text-intensive tasks, our findings reveal significant concerns about trustworthiness, transparency, and hallucination. Some observations are likely time-sensitive (e.g., specific product integrations or grounding features). However, the time-stable issues raised by participants—verification overhead, responsibility for outputs, and where LLM suggestions can safely enter workflows—are shaped by organizational process and remain relevant even as tools evolve. We discuss key implications, contextualize findings within existing research, and provide recommendations for deploying LLM-driven security solutions. \textcolor{black}{Several of our findings are consistent with broader literature on LLM reliability and AI-assisted workflows, particularly concerns around hallucinations, prompt sensitivity, verification burden, and continued reliance on human oversight. What appears especially salient in SOC settings, however, is the combination of these issues with operational time pressure, the need to trace claims back to logs, rules, or threat-intelligence evidence, and governance constraints on how model-generated content may enter detection and response pipelines \cite{badva2024unveiling, saha2025expert, maxam2024interview}. These factors help explain why participants were more comfortable using LLMs for quickly verifiable, draft-oriented tasks than for investigative reasoning or higher-stakes decision support.}

\textbf{How to use the artifacts in practice.}
The artifacts support three decision points: engineers/leads use the \emph{Constraint Matrix} to choose safe SOC use cases, management uses the \emph{maturity rubric} to gate deployment based on guardrails, and analysts use the \emph{failure-mode taxonomy} to guide verification and escalation of LLM outputs during investigations.

\subsection{Addressing Practical Barriers: Data, Integration, and Organizational Hurdles}
\textcolor{black}{\textbf{Observed practice:} Participants described that adoption is often constrained by data governance, integration overhead, resourcing, and deployment costs, which can stall pilots before full operational integration.
\textbf{Expressed need:} They want clearer compliance pathways and reliable internal pipelines that make SOC-specific grounding feasible without relying on informal, case-by-case exceptions.
\textbf{We propose:} Treat LLM adoption as a data-and-governance program, not only a model-selection decision. Near-term efforts should prioritize building auditable data pipelines that capture and govern logs, threat intelligence feeds, and analyst annotations under explicit access and usage rules, in close coordination with legal and compliance stakeholders. Related decentralized log-analysis research illustrates one potential architecture for addressing this tension. Federated LogTracer extracts contextual relationships and identifies malicious logs while retaining sensitive telemetry locally rather than consolidating it on a central analysis server \cite{rabieinejad2026united}. Although our interview study does not evaluate this architecture, it demonstrates how SOC-specific contextualization can be pursued under privacy and data-locality constraints. For organizations exploring model localization, we propose evaluating iterative refinement approaches (e.g., human-in-the-loop updating/active learning) \cite{lu2023human} and efficiency-oriented fine-tuning strategies (e.g., fractional fine-tuning) \cite{hubotter2024efficiently} as options to reduce operational cost and privacy friction, while remaining explicit that our interview study does not assess their effectiveness or cost-benefit in situ. Finally, we propose that executive sponsors and SOC leadership define upfront the data rights, privacy policies, and infrastructure commitments required for pilots to transition into sustained operational tools.}

\subsection{Mitigating Hallucinations in High-Stakes Security Tasks}

\textcolor{black}{\textbf{Observed practice:} Participants primarily managed incorrect outputs through human-in-the-loop verification and cross-referencing against logs, indicators, and internal knowledge.
\textbf{Expressed need:} They want mitigations that reduce verification drag while making it clearer when an output is grounded versus speculative.
\textbf{We propose:} Evaluate layered mitigation designs that align with these needs: (i) grounding outputs using retrieval over curated internal/CTI sources (e.g., RAG) as a way to attach provenance to claims \cite{Lewis2020RAG}; (ii) redundancy or cross-checking across models or conventional systems to flag inconsistencies \cite{zhang2023sac3}; and (iii) monitoring approaches that periodically re-validate high-impact outputs as context and threat information shift \cite{illakiya2024ai}. These are proposed design directions motivated by participant reports; our study does not empirically validate their effectiveness.}


\subsection{Designing Actionable Explanations and Workflow Integration}


\textcolor{black}{\textbf{Observed practice:} Participants preferred concise outputs that could be quickly checked against familiar SOC artifacts (e.g., detection rules, known indicators, and raw logs), and were less receptive to long narratives when verification was unclear.\\
\textbf{Expressed need:} They want explanations that are operationally actionable and easy to validate under time pressure.\\
\textbf{We propose:} Design interfaces around verification-first explanations: provide a short triage-ready summary plus a drill-down view that links claims to concrete evidence (e.g., referenced log lines, query results, or CTI items) and maps key elements to recognized frameworks (e.g., MITRE ATT\&CK) when applicable. We also propose lightweight feedback channels (e.g., flag/annotate outputs and capture “what evidence was used”) so organizations can iteratively align outputs with local workflows without assuming that feedback automatically improves reliability.}

\subsection{Operationalizing Human Factors: Codifying Analyst Resourcefulness}
\textcolor{black}{\textbf{Observed practice:} Participants described a set of operator-led guardrails, evidence-only prompting, checklist-driven triage, cross-checks against SOC systems, structured outputs for validation, and dual-control for higher-impact actions, as practical ways to bound incorrect outputs without removing human judgment.\\
\textbf{Expressed need:} They want these practices to be repeatable and auditable, rather than dependent on individual vigilance.\\
\textbf{We propose:} Codify these guardrails into runbooks and tooling: (i) adopt an evidence-only policy for prompts and responses; (ii) maintain a curated prompt/checklist library for common triage and drafting tasks; (iii) require structured outputs with schema validation and explicit provenance fields (e.g., referenced log IDs, CTI citations); (iv) apply dual-control review for high-impact actions or changes derived from LLM suggestions; and (v) instrument lightweight verification logs that record what was checked and on which evidence \cite{hokamp-liu-2017-lexically, lewis2020retrieval}. These are process and tooling directions grounded in participant-described practice; we do not claim they eliminate failure modes, particularly under novelty or time pressure.}

\subsection{The Impact of Organizational Maturity on LLM Integration}
\textcolor{black}{\textbf{Observed practice:} Participants described substantial heterogeneity: smaller teams often rely on commodity interfaces for general assistance, while larger enterprises/MSSPs described more structured (but still experimental) integration within SIEM/XDR toolchains under stricter governance. This pattern is consistent with CISO-level evidence showing that enterprise GenAI readiness is generally stronger in upstream governance than in runtime assurance, where limited telemetry and operational procedures leave detection largely human-in-the-loop \cite{rabieinejad2026orgwide}.\\
\textbf{Expressed need:} Organizations want solutions that match their maturity constraints, low-maturity SOCs need usable guardrails without specialized AI expertise, while higher-maturity SOCs need auditability and policy alignment.\\
\textbf{We propose:} Develop maturity-tailored configurations aligned with the rubric (Table~\ref{tab:hallucination-mitigation-maturity}). For lower-maturity environments, prioritize safe defaults and bounded workflows (e.g., evidence-only prompting, structured templates, and clear escalation boundaries) that reduce reliance on ad hoc judgment. For higher-maturity environments, prioritize API-driven integrations that support governance controls (e.g., access restrictions, audit logging, and approval gates) and allow controlled experimentation under compliance constraints. We also propose SOC-facing AI maturity benchmarks that help organizations self-assess readiness for higher-autonomy use (e.g., expected verification overhead and accountability requirements) without implying that any specific configuration is validated in this study.}

\section{Conclusion}\label{s8}
\textcolor{black}{Our study of 20 security practitioners suggests that LLMs are being cautiously integrated into SOC workflows primarily as supplementary tools. Participants report perceived time savings for text-heavy tasks (e.g., summarization and drafting), typically conditional on substantial verification. However, participants describe limited trust in LLMs for high-stakes security decisions due to unreliable outputs and opaque reasoning. Participants also seldom described standardized, documented mitigation procedures, relying instead on ad-hoc verification norms and continuous human oversight.
Overall, these accounts point to a hybrid reality where LLM outputs serve as preliminary leads or drafts that must be corroborated using existing SOC tooling and expert judgment. Using a maturity rubric applied at the participant level, we find readiness is uneven and often reflects informal practices rather than governed processes. Future work should examine how vendor-hosted deployment models shape privacy, data-residency, and auditability constraints in production SOC environments.}

\appendix
\renewcommand{\thesection}{\appendixname~\Alph{section}}
\renewcommand{\thesubsection}{\Alph{section}.\arabic{subsection}}

\appendix
\renewcommand{\thesection}{\Alph{section}}

\titleformat{\section}
  {\normalfont\bfseries}
  {APPENDIX \thesection.}
  {0.5em}
  {}

\section{Ethical Considerations}
This study received institutional ethics approval (protocol details withheld for double-blind review). All participants provided informed consent and spent approximately 35 minutes in total.
Relevant stakeholders include participants, their organizations, the broader SOC community and end users, vendors, and potential attackers. Key risks include re-identification, disclosure of operationally sensitive practices, reputational or organizational harm, misuse of findings to evade detection, and misinterpretation that amplifies analyst displacement narratives.
We mitigated these risks through data minimization and conservative reporting: no PII was retained in the research dataset; individual and organization names were not kept beyond scheduling; scheduling and remuneration emails were stored separately and deleted after compensation; transcripts were anonymized and stored securely. In reporting, we avoid operationally enabling details, paraphrase rather than quote when sensitivity or identification risk is non-trivial, and release only decision-guiding artifacts (taxonomy, rubric, constraint matrix) rather than operationally prescriptive instructions.
Overall, we judge that the benefits of workflow-grounded guidance for safer, more verifiable LLM use in SOCs outweigh residual risks, which are further reduced by not releasing identifying details, operationally sensitive specifics, or raw transcripts. \textcolor{black}{More explicitly, the stakeholders in this work include participants, their organizations, vendors, the broader SOC community, downstream users of SOC outputs, and potential attackers. The main publication risk is not only re-identification, but also that practitioner accounts could be overgeneralized in ways that normalize premature or weakly governed deployment. We therefore frame the contribution as decision-guiding rather than deployment-prescriptive, and we judge publication to be warranted because practitioner-grounded evidence is needed to support safer, more governable LLM adoption in SOC environments.}

\section{Open Science}
\textcolor{black}{To support transparency during double-blind review, we provide an anonymized materials package at \url{https://anonymous.4open.science/r/llm-soc-usability-artifacts-52E4/} (interview guide, codebook, and participant-facing documents). Materials are scrubbed of institutional identifiers and author metadata to preserve double-blind review. Due to ethics and re-identification risk in operational security settings, we do not release raw audio/transcripts, organizational metadata, participant-level coding vectors, or any verbatim transcript excerpts; exemplar snippets in the paper are paraphrased/abstracted. We therefore emphasize process auditability via shared instruments/codebook and make derived outputs explicit and checkable within the paper.}

\bibliographystyle{plainurl}
\bibliography{software}

\section{Interview Protocol and Codebook}\label{a1}
This appendix presents the interview guide and the codebook used for thematic analysis, including frequencies and descriptions. Tables~\ref{t2} and~\ref{t3} summarize the analytical framework and recurring themes.

\subsection{Semi-Structured Questions}\label{p1}
The following prompts guided all interviews; follow-ups were used to probe for detail. Questions align with the dimensions in Tables~\ref{t2} and~\ref{t3}.

\begin{enumerate}
    \item Can you describe your current role and responsibilities within the SOC?
    \item Have you had any experience using LLM-based tools in your SOC?
    \item What are the primary benefits you have observed from using LLMs in your SOC?
    \item What challenges or limitations have you encountered when using LLMs?
    \item How do LLMs compare to traditional rule-based systems in performance and usability?
    \item How adaptable do you find LLMs to different security environments and threats?
    \item How often do you update or retrain your LLM models?
    \item Have you encountered hallucinations (incorrect or misleading information)?
    \item Are you familiar with hallucination mitigation techniques?
    \item How intuitive do you find the user interface of LLM-based tools?
    \item How do you provide feedback to LLM-based tools, and how is it incorporated?
\end{enumerate}

\subsection{Interview Codebook}\label{p2}
The codebook, developed from analyses of participant interviews, appears in Tables~\ref{t2} and~\ref{t3}.
\begin{table*}[t]
\centering
\scriptsize
\caption{Interview Codebook — Coded Descriptions of Interview Themes}
\label{t2}
\resizebox{1.5\columnwidth}{!}{%
\begin{tabular}{p{3.7cm} p{4.0cm} p{1.0cm} p{7.0cm}}
\toprule
\textbf{Primary Code} & \textbf{Subcode} & \textbf{Freq.} & \textbf{Description} \\
\midrule
\textbf{Domain} & IT Cybersecurity & 11/20 & Organizations engaged in threat hunting, rule formulation, or comprehensive SOC operations. \\
 & Network Solutions & 2/20 & Entities specializing in network infrastructure and related security services. \\
 & Legal & 1/20 & Organizations operating within legal frameworks or law practices. \\
 & Public Sector & 1/20 & Institutions in the public or governmental domain focused on governance and risk management. \\
 & Telecommunication & 1/20 & Organizations in the telecommunications industry delivering or securing network services. \\
 & Sports/Entertainment & 1/20 & Entities within the sports or entertainment sectors that maintain dedicated cybersecurity functions. \\
 & Finance & 3/20 & Organizations in the financial sector including banks and financial institutions. \\
 & Food Retailing & 1/20 & Organizations in the retail food industry with in-house cybersecurity operations. \\
\midrule
\textbf{Technical Task} & Threat Hunting & 6/20 & Systematic search for indicators of compromise or anomalous patterns within logs or network traffic. \\
 & SOC Analysis & 3/20 & First-line triage and investigation of security alerts. \\
 & SOC Leadership & 1/20 & Oversight and coordination of SOC analysts and daily operations. \\
 & Technical Account Management & 1/20 & Provision of technical support and consultation regarding security services. \\
 & Information Security Analysis & 1/20 & Routine evaluation of potential security incidents. \\
 & Anomaly Detection & 1/20 & Identification of deviations from normal operational behavior. \\
 & Threat Intelligence & 3/20 & Collection and analysis of adversary tactics, techniques, and procedures. \\
 & Detection Engineering & 4/20 & Development and maintenance of detection rules and logic. \\
 & Cyber Operations Management & 3/20 & Strategic management of cybersecurity teams and incident response protocols. \\
 & Threat Intelligence Leadership & 1/20 & Leadership roles that combine operational and research-based threat analysis. \\
 & Incident Response / DFIR & 7/20 & Execution of incident response and digital forensics procedures. \\
 & Security Platform Management & 2/20 & Administration of security tools such as firewalls, endpoint detection, and cloud security systems. \\
 & Security Advisory & 1/20 & Provision of expert guidance on cybersecurity matters. \\
 & Model Tuning & 1/20 & Customization and refinement of detection models for specific environments. \\
 & Dashboard Review & 1/20 & Analysis of aggregated security data and performance dashboards. \\
 & Security Research & 3/20 & Investigation of emerging threats and advanced security methodologies. \\
 & Playbook Development & 1/20 & Creation and revision of standardized incident response procedures. \\
 & Incident Command & 1/20 & Coordination of security incident management. \\
 & Solution Architecture & 1/20 & Design and implementation of security systems and solutions. \\
 & CCR Team Management & 1/20 & Supervision of teams focused on containment and threat mitigation. \\
 & Data Science in Security & 1/20 & Application of data-driven models and AI techniques in cybersecurity. \\
\midrule
\textbf{Non-Technical Task} & Management/Coordination & 2/20 & Oversight of security teams, budget management, and vendor relations. \\
 & Reporting/Governance & 2/20 & Preparation of reports, policies, and ensuring regulatory compliance. \\
 & Stakeholder Interaction & 2/20 & Communication with internal and external stakeholders regarding security issues. \\
 & Security Training & 2/20 & Delivery of cybersecurity awareness and training programs. \\
\midrule
\textbf{Task Output Recipient} & Internal Security Team & 12/20 & SOC analysts and internal staff who receive and act upon security alerts. \\
 & Site Reliability Engineering & 1/20 & Technical teams responsible for maintaining large-scale systems. \\
 & Executive Management & 1/20 & Senior leadership reviewing high-level security summaries. \\
 & Platform Engineering & 1/20 & Engineering teams responsible for system infrastructure and platform maintenance. \\
 & Datacenter Administration & 1/20 & Management of data center operations. \\
 & R\&D Team & 1/20 & Research and development groups focused on security innovation. \\
 & Threat/Triage Response & 1/20 & Specialized teams for threat identification and incident response. \\
 & Clients & 4/20 & External customers receiving security solutions and advisories. \\
 & Senior Management/Directors & 6/20 & Organizational leadership involved in strategic security decisions. \\
 & Vendors/MSSP & 4/20 & Third-party security providers or managed service partners. \\
\midrule
\textbf{Current Tools} & Web Security Solutions & 12/20 & Next-generation firewalls, endpoint detection and response systems, and cloud security platforms. \\
 & HDR Systems & 1/20 & Host detection and response solutions. \\
 & Security Gateways & 1/20 & Perimeter devices controlling network data flow. \\
 & ETL Pipelines & 1/20 & Systems for extracting, transforming, and loading security logs. \\
 & Rule Engines & 1/20 & Platforms utilizing manually crafted rules or signatures. \\
 & Large-Scale Search Platforms & 1/20 & Tools designed for indexing and searching extensive datasets. \\
 & Prisma & 1/20 & Commercial security solutions referenced for specific functionalities. \\
 & SIEM/XDR Tools & 7/20 & Integrated platforms for log analysis and threat detection (e.g., CrowdStrike, Microsoft Defender). \\
 & DLP Solutions & 2/20 & Systems for preventing data leakage. \\
 & Vulnerability Management & 1/20 & Tools such as Qualys for vulnerability assessment. \\
 & Cloud Databases & 1/20 & Databases employed for storage of security data. \\
 & Cisco XDR & 1/20 & Cisco's extended detection and response suite. \\
 & AI-Based Security Tools & 2/20 & Security solutions incorporating artificial intelligence techniques. \\
 & Forensic Tools & 4/20 & Software for digital forensics and incident investigation. \\
 & Federated Search & 1/20 & Unified search systems for aggregating security data. \\
 & Threat Intelligence Platforms & 5/20 & Commercial or public platforms providing curated threat intelligence. \\
 & Anomaly Detection ML Tools & 2/20 & Machine learning solutions for identifying anomalous patterns. \\
 & Custom Security Tools & 1/20 & In-house developed tools for security analysis. \\
 & LLM-Based Security Tools & 2/20 & Solutions that leverage LLMs for security applications. \\
 & Security Awareness Tools & 1/20 & Systems for managing and organizing security training programs. \\
\midrule
\textbf{Level of LLM Usage} & Continuous Use & 11/20 & Regular and integrated use of LLM-based tools in daily operations. \\
 & Occasional/Individual Use & 6/20 & Sporadic use by individual analysts. \\
 & Not Utilized & 3/20 & Organizations that do not employ in-house LLM-based solutions. \\
\midrule
\textbf{Benefits of LLM} & Efficiency/Time-Saving & 13/20 & Automates repetitive tasks, thereby increasing operational efficiency. \\
 & Task Automation & 5/20 & Enables partial or complete automation in various security processes. \\
 & Support for Lateral Thinking & 2/20 & Facilitates creative approaches and alternative analytical perspectives. \\
 & Workload Reduction & 3/20 & Alleviates routine operational burdens. \\
 & Enhanced Documentation & 3/20 & Improves the quality and consistency of security documentation. \\
 & Data Analysis & 1/20 & Assists in processing and analyzing large datasets. \\
 & Simplification & 1/20 & Translates complex technical information into more accessible language. \\
 & Decoding Obfuscated Code & 1/20 & Aids in the interpretation of ambiguous or complex scripts. \\
 & Process Mapping & 1/20 & Summarizes and clarifies complex operational processes. \\
 & Summarization of Texts & 4/20 & Condenses extensive logs and reports into concise summaries. \\
 & Standardized Communication & 1/20 & Ensures uniformity in internal communications. \\
 & Knowledge Augmentation & 2/20 & Provides rapid access to best practices and reference information. \\
\midrule
\end{tabular}
}
\end{table*}

\begin{table*}[t]
\centering
\scriptsize
\caption{Interview Codebook — Coded Descriptions of Interview Themes}
\label{t3}
     \resizebox{1.5\columnwidth}{!}{%
\begin{tabular}{p{3.7cm} p{4.0cm} p{1.0cm} p{7.0cm}}
\toprule
\textbf{Primary Code} & \textbf{Subcode} & \textbf{Freq.} & \textbf{Description} \\
\midrule
\textbf{Challenges \& Limitations of LLM} & Hallucination Risk & 11/20 & Potential for generating plausible but inaccurate responses. \\
 & Overly General Output & 1/20 & Insufficient detail or task-specific precision. \\
 & Lack of Formalized Prompt Guidelines & 4/20 & Absence of standardized prompt design protocols. \\
 & Inability to Maintain Context & 1/20 & Difficulties in sustaining topic focus during interactions. \\
 & Variability in Responses & 1/20 & Inconsistent outputs from identical inputs. \\
 & Misclassification Errors & 6/20 & Occurrence of false positives and false negatives. \\
 & Requirement for High-Quality Data & 2/20 & Necessity for curated and standardized data inputs. \\
 & Inadequate Contextual Coverage & 2/20 & Omission of critical contextual information. \\
 & Limitations in Context Length & 1/20 & Challenges with processing multi-step or voluminous data. \\
 & Verification Overhead & 4/20 & Need for manual cross-verification of model outputs. \\
 & Limited Retraining Frequency & 1/20 & Infrequent updates leading to potential model staleness. \\
 & Insufficient LLM Expertise & 2/20 & Deficit in specialized skills for managing LLMs. \\
 & High Time Investment & 1/20 & Significant effort required for setup and ongoing management. \\
 & Resource Limitations & 3/20 & Budgetary or computational constraints affecting model fine-tuning. \\
 & Suboptimal for Incident Response & 1/20 & Limited efficacy in real-time incident response scenarios. \\
 & Latency Issues & 2/20 & Delays in model inference impacting time-sensitive tasks. \\
 & Opaque Decision Processes & 1/20 & Lack of transparency in the model’s internal reasoning. \\
\midrule
\textbf{Factor to Choose LLM} & Trustworthiness & 3/20 & Confidence in the reliability of the LLM’s outputs. \\
 & Task Suitability & 2/20 & Preference for language-centric applications. \\
 & Available Support & 1/20 & Access to technical support or vendor assistance. \\
 & Ease of Use & 3/20 & Low adoption friction and minimal user effort. \\
 & Performance Evaluations & 1/20 & Positive industry benchmarks and reviews. \\
 & Time Efficiency & 6/20 & Demonstrated improvements in operational speed. \\
 & Scalability & 2/20 & Ability to manage increased data volumes or incident frequency. \\
 & Accuracy & 5/20 & Proximity to ground truth with minimal misclassifications. \\
\midrule
\textbf{Adaptability} & General Adaptability & 3/20 & Effective performance with standard threat patterns and data. \\
 & High Adaptability & 4/20 & Superior flexibility when provided with high-quality data and context. \\
 & Limitations for Novel Attacks & 3/20 & Reduced efficacy with emerging or obscure threat vectors. \\
 & Inapplicability for Statistical Analysis & 1/20 & Unsuitability for tasks requiring precise numerical computations. \\
 & Proficient in Data Analysis & 3/20 & Capable of processing logs, IPs, and other structured data for threat detection. \\
 & Limited for High-Certainty Tasks & 1/20 & May require supplementary tools for exact computations. \\
 & Infrastructure Constraints & 4/20 & Usage limited by compliance or environmental factors. \\
 & Dependence on Data Recency & 1/20 & Performance contingent on the freshness of the input data. \\
 & High Alignment Effort & 7/20 & Significant organizational effort needed for policy and procedural alignment. \\
 & Reduced Effort Compared to Traditional ML & 2/20 & Lower implementation burden when using pre-trained models. \\
 & Uncertain Adaptability & 1/20 & Potential applicability in niche domains such as digital forensics. \\
\midrule
\textbf{LLM Application at Work} & Investigative Assistance & 3/20 & Provides guidance and preliminary steps for incident investigations. \\
 & Structuring Unstructured Data & 1/20 & Organizes and clarifies unstructured textual information. \\
 & Script/Regex Generation & -- & Automates creation of small code snippets or expressions. \\
 & Drafting Investigation Steps & 5/20 & Generates initial checklists and procedures for incident response. \\
 & Documentation & 4/20 & Summarizes and standardizes security documentation. \\
 & Transactional Risk Scoring & 1/20 & Integrates vulnerability and risk data for assessment. \\
 & Log Analysis & 6/20 & Efficient summarization and analysis of large-scale log data. \\
 & IoC Extraction & 3/20 & Identifies and extracts indicators of compromise. \\
 & Malware Detection Support & 2/20 & Assists in the classification or analysis of potential malware. \\
 & Gap Analysis & 1/20 & Evaluates the comprehensiveness of detection strategies. \\
 & Risk Assessment Insights & 1/20 & Provides summarized risk evaluations for security assets. \\
 & Automated Pentesting & 1/20 & Generates preliminary exploit paths or test scenarios. \\
 & Detection Suggestions & 3/20 & Proposes modifications or enhancements to detection logic. \\
 & Communication Enhancement & 1/20 & Improves clarity in written reports and communications. \\
 & Alert Summarization & 2/20 & Aggregates and condenses multiple alerts into coherent summaries. \\
 & BI Integration & 1/20 & Integrates LLM outputs with business intelligence dashboards. \\
 & Contextual Information Provision & 2/20 & Supplies additional context to support investigative decision-making. \\
 & Specialized BI Integration & 1/20 & Facilitates import of summarized data into BI tools. \\
 & Custom Detection Evaluation & 3/20 & Assesses and validates custom detection filters. \\
\midrule
\textbf{Comparison to Rule-Based} & Superior for Language-Heavy Tasks & 2/20 & Excels in processing and analyzing unstructured textual data. \\
 & Time Efficiency & 3/20 & Achieves faster processing with reduced manual intervention. \\
 & Opaque Model Operation & 1/20 & Operates as a black box relative to transparent rule-based systems. \\
 & Enhanced Pattern Recognition & 1/20 & Capable of detecting subtle patterns beyond predefined rules. \\
 & Complexity in Tuning & 1/20 & More challenging to modify compared to manually adjustable rules. \\
 & Reduced Coding Requirements & 1/20 & Minimizes the need for extensive programming in certain tasks. \\
 & Flexible Analytical Perspectives & 2/20 & Evaluates data from multiple viewpoints beyond fixed patterns. \\
 & Accessibility Across Experience Levels & 1/20 & Usable by personnel with varying levels of expertise. \\
 & Minimal AI Reliance & 3/20 & Some users prefer limited reliance on AI compared to rule-based methods. \\
\midrule
\textbf{Effectiveness Factors} & Not Evaluated & 2/20 & Absence of formal evaluation metrics. \\
 & Feedback Mechanism & 1/20 & Incorporates user feedback for continuous model improvement. \\
 & Customized Benchmarking & 1/20 & Utilizes tailored test scenarios for performance evaluation. \\
 & Consistency Verification & 1/20 & Ensures stable outputs across repeated queries. \\
 & Algorithmic Trustworthiness & 1/20 & Subjective evaluation of model reliability. \\
 & Operational Time Efficiency & 1/20 & Reduction in incident response times. \\
 & Industry Feedback & 2/20 & Based on external reviews and best practices. \\
 & Standard ML Metrics & 7/20 & Employs precision, recall, and related metrics. \\
 & Business Impact Metrics & 1/20 & Evaluates performance based on customer satisfaction. \\
 & Contextual Enrichment & 1/20 & Provides supplementary context to aid decision-making. \\
 & Manual Validation & 12/20 & Reliance on expert review to confirm LLM outputs. \\
\midrule
\textbf{LLM Explanations \& UI} & Reliance on Vendor Interface & 4/20 & Utilizes the vendor-provided user interface for interaction. \\
 & Open Source GPT Interface & 7/20 & Employs a chat-based interface with transparency features. \\
 & Scope for Improvement & 4/20 & Indicates potential for enhanced integration and error flagging. \\
 & Lack of Transparency & 1/20 & Perceived opacity in the model’s decision-making process. \\
 & Indeterminate UI Evaluation & 3/20 & Uncertainty in assessing interface clarity. \\
\midrule
\textbf{Feedback in UI} & In-Tool Feedback Mechanisms & 4/20 & Features for user corrections and annotations within the interface. \\
 & Absence of Formal Feedback Submission & 1/20 & No direct feedback channels established with model developers. \\
 & Alternative Feedback Channels & 2/20 & Utilization of external communication methods (e.g., Slack). \\
 & Lack of Structured Feedback & 4/20 & Absence of a systematic approach for gathering user input. \\
 & Regular Feedback Meetings & 5/20 & Periodic review sessions with vendors or partners. \\
\midrule
\textbf{Handling Hallucinations} & Absence of Formal Methodology & 8/20 & Reliance on ad hoc approaches for mitigating fabricated outputs. \\
 & Prompt Refinement & 3/20 & Modification of prompts to reduce model guesswork. \\
 & Human Oversight & 10/20 & Continuous expert review of outputs to ensure accuracy. \\
 & Use of Confidence Metrics & 1/20 & Incorporation of confidence scoring for validation purposes. \\
 & Self-Correction Mechanisms & 2/20 & Implementation of internal re-evaluation of outputs. \\
 & Retrieval-Augmented Generation & 1/20 & Grounding outputs using verified external sources. \\
 & Standard Reference Alignment & 1/20 & Labeling outputs with established standards for verification. \\
 & Iterative Output Consistency & 1/20 & Repetition of processing to confirm stable results. \\
 & Cross-Model Comparison & 1/20 & Comparison of outputs from multiple models for consistency. \\
 & Hard Validation Tools & 1/20 & Use of specialized tools for output verification. \\
 & Cross-Referencing Techniques & 1/20 & Matching outputs against known data sources. \\
 & Schema-Based Validation & 1/20 & Validation of outputs against predefined schemas. \\
\midrule
\textbf{Frequency of LLM Retraining} & Uncertain & 9/20 & Frequency of model updates remains unclear. \\
 & No Retraining & 2/20 & Use of pre-trained models without further retraining. \\
 & Periodic Retraining (1--3 Months) & 2/20 & Regular updates on a short-term basis. \\
 & Conditional Retraining & 2/20 & Updates deployed based on performance improvements. \\
 & Infrequent Retraining & 1/20 & Occasional model updates. \\
 & Event-Driven Updates & 3/20 & Updates triggered by significant incidents or anomalies. \\
\bottomrule
\end{tabular}
}
\end{table*}

\section{Participants' Demographic and Professional Information}\label{b1}
This section details the profiles of the 20 participants, including their roles, experience levels, and primary security domains.

\begin{table*}[t]
    \centering
    \caption{Overview of Participants' Demographic and Professional Information}
    \label{t1}
    \resizebox{\textwidth}{!}{%
        \begin{tabular}{llllllll}
            \toprule
            ID  & Job Role                                  & Industry                     & Experience (years) & Security Domain       & Detection Method        & AI Expertise             & Location       \\
            \midrule
            P01 & Security Analyst                          & Cybersecurity                & 3                  & Defensive             & Both                   & Practical Application    & Ontario, CA    \\
            P02 & Security Manager                          & Cybersecurity                & 20                 & Offensive \& Defensive & Rule-based             & Conceptual Understanding & Ontario, CA    \\
            P03 & Director, Cybersecurity Operations        & Food Retail                  & 17                 & Offensive \& Defensive & Both                   & Practical Application    & Ontario, CA    \\
            P04 & Threat Response \& Detection Engineer     & Cybersecurity                & 14                 & Defensive             & ML                     & Practical Application    & Ontario, CA    \\
            P05 & Solution Architect                        & Network Solutions            & 22                 & Defensive             & Rule-based             & Methodological Knowledge & Ontario, CA    \\
            P06 & Security Analyst                          & Finance                      & 7                  & Defensive             & Rule-based             & Practical Application    & Ontario, CA    \\
            P07 & Data Scientist                            & Cybersecurity                & 6                  & Defensive             & ML                     & Practical Application    & Ontario, CA    \\
            P08 & Security Researcher                       & Cybersecurity                & 15                 & Defensive             & Both                   & Practical Application    & Ontario, CA    \\
            P09 & Security Analyst                          & Cybersecurity                & 4                  & Defensive             & Both                   & Practical Application    & Ontario, CA    \\
            P10 & Security Analyst                          & Telecommunications           & 3                  & Defensive             & Both                   & Practical Application    & Ontario, CA    \\
            P11 & SOC Manager                               & Finance                      & 8                  & Offensive \& Defensive & Both                   & Practical Application    & Ontario, CA    \\
            P12 & SOC Manager                               & IT                           & 25                 & Offensive \& Defensive & ML                     & Practical Application    & Ontario, CA    \\
            P13 & Security Analyst                          & IT                           & 3                  & Defensive             & Rule-based             & Methodological Knowledge & Ontario, CA    \\
            P14 & Technical Lead                            & IT                           & 10                 & Defensive             & Both                   & Practical Application    & Ontario, CA    \\
            P15 & Cyber Crisis Response (CCR) Manager       & Finance                      & 8                  & Defensive             & Rule-based             & Conceptual Understanding & Ontario, CA    \\
            P16 & Security Analyst                          & Cybersecurity                & 1.5                & Defensive             & Rule-based             & Conceptual Understanding & Ontario, CA    \\
            P17 & Technical Account Manager                 & Computer \& Network Security & 5                  & Defensive             & Both                   & Practical Application    & California, USA\\
            P18 & Security Analyst                          & Legal                        & 4                  & Defensive             & Rule-based             & Conceptual Understanding & Ontario, CA    \\
            P19 & Director, Threat Intelligence             & Cybersecurity                & 10                 & Defensive             & Both                   & Practical Application    & Ontario, CA    \\
            P20 & SOC Team Lead                             & IT                           & 5                  & Defensive             & Both                   & Practical Application    & Ontario, CA    \\
            \bottomrule
        \end{tabular}
    }
\end{table*}

\section{Thematic Frameworks and Taxonomy}\label{c1}
This appendix consolidates the paper’s derived artifacts for auditability. We include (i) the failure-mode taxonomy with inclusion/exclusion rules (Table~\ref{tab:failure-mode-taxonomy}), (ii) hallucination-risk outcome types and reported mitigations (Table~\ref{tab:hallucination-risk-model}), (iii) the participant-reported maturity rubric with brief exemplars (Table~\ref{tab:hallucination-mitigation-maturity}), and (Iv) a role-stratified summary of risks and implications (Table~\ref{tab:stratified-role-analysis}).

\begin{table*}[t]
\centering
\color{black}
\footnotesize
\caption{Transcript-derived taxonomy of LLM output failure modes in SOC workflows, with inclusion/exclusion criteria and severity mapping (consequence-if-acted-upon).}
\label{tab:failure-mode-taxonomy}
\begin{tabular}{p{0.12\textwidth} p{0.27\textwidth} p{0.20\textwidth} p{0.08\textwidth} p{0.27\textwidth}}
\hline
\textbf{Class} & \textbf{Inclusion (what it is)} & \textbf{Exclude (boundary rule)} & \textbf{Severity} & \textbf{Representative excerpt(s)} \\
\hline
FM1: Fabricated security facts & Output asserts specific ``facts'' (e.g., incident details, IOCs, TTPs) that are not grounded in provided logs/CTI/internal knowledge bases. & Clearly labeled hypotheses (``might be...'') \emph{with explicit evidence pointers} or requests for additional data. & High & P06: ``Most of us know LLMs can make things up, but we’re not sure how to systematically prevent it... hope no one junior takes it at face value.'' \\
\hline
FM2: Invalid detection logic / syntax errors & Output produces nonexistent operators, invalid query/rule syntax, or wrong rule constructs that appear plausible and may be copied into tooling. & Pure formatting/style suggestions that do not change semantics or executable behavior. & High & P04: an LLM ``invented operators'' in detection rule syntax, requiring extensive manual correction. \\
\hline
FM3: Misleading investigative direction & Output recommends incorrect next steps, prioritization, or triage checklists that steer analysts away from evidence or waste effort. & Benign brainstorming explicitly separated from action steps (e.g., alternatives framed as optional). & Medium & P09: ``We treat the LLM output as a lead, not as a conclusion.''  \\
\hline
FM4: Incorrect interpretation yielding FP/FN & Output misinterprets suspicious data and produces false positives/negatives in labeling, explanation, or triage conclusions. & Cases where the model refuses to conclude and asks for more evidence/context. & High & P03, P11, P19: Participants reported persistent interpretation errors where models mislabeled benign patterns as malicious. \\
\hline
FM5: Opaque / unverifiable claims & Output provides an answer without traceable rationale (no ``why,'' no linkage to rule triggers/log lines), increasing verification load and slowing response. & Outputs that explicitly cite evidence (log lines, query results, KB entries) or provide reproducible reasoning steps. & Medium & P05: ``With rule-based systems, at least we know which rule triggered. The LLM just gives an answer, and I’m not always sure why.'' \\
\hline
FM6: Guessing under novelty / sparse intel & Under novel threats (e.g., fresh zero-days) or minimal context, output fills gaps by inference and can become off-base or overconfident. & Outputs that explicitly decline to speculate and request specialized threat intelligence or additional telemetry. & Medium & P19: ``when it’s a fresh zero-day or we have minimal intel, the LLM is forced to guess, and sometimes that guess is off-base.'' \\
\hline
FM7: Verification overhead eroding efficiency & The cost of checking output meaningfully cancels time saved, particularly under time pressure or high-stakes reporting. & Routine quick checks in low-stakes tasks where verification is minimal and not burdensome. & Low & P04: ``We’ll save an hour drafting rules but spend half an hour verifying every line.''  \\
\hline
\end{tabular}
\end{table*}

\begin{table*}[t]
\centering
\color{black}
\footnotesize
\caption{Hallucination-risk outcome types observable in our interview corpus.}
\label{tab:hallucination-risk-model}
\begin{tabular}{p{0.20\textwidth} p{0.30\textwidth} p{0.18\textwidth} p{0.26\textwidth}}
\hline
\textbf{Outcome type (failure)} & \textbf{Operational consequence (what “risk” means)} & \textbf{Example evidence (participants)} & \textbf{Mitigation practices reported} \\
\hline
Fabricated facts / entities & Plausible misinformation (e.g., fictitious incident narratives, nonexistent IOCs/TTPs) increases verification burden and may misdirect investigation if treated as a lead. & P06 (fabrication risk; lack of systematic prevention); Section 6.1 examples of fictitious IOCs/TTPs. & Evidence-based verification against logs/CTI; treat output as a lead not a conclusion (P09); strict human review and sign-off (P08). \\
\hline
Schema-invalid / tool-invalid artifacts & Invalid detection content (e.g., rule syntax) causes rework, delays, and can introduce errors into SOC tooling pipelines. & P04 (“invented operators” in detection rule syntax). & Structured output constraints; schema validation; dual-control review for rule changes; human correction loops. \\
\hline
Misleading procedural recommendations & Incorrect “what to do next” guidance (e.g., first-pass triage steps) can waste time, increase cognitive load, or propagate error into the workflow. & P13 (LLM drafted “first-pass triage steps”); participants reporting need for review due to inaccuracies/logical errors. & Prompt constraints to avoid assumptions (P07); human oversight as final authority (P08); cross-checking via redundancy (e.g., multi-model comparison, P17). \\
\hline
\end{tabular}
\end{table*}

\begin{table*}[t]
\centering
\footnotesize
\color{black}
\caption{Participant-reported maturity rubric for LLM mitigation in SOC workflows ($N{=}20$), assigned from interview transcripts (Section~6.3). \textit{Not applicable } = no workplace LLM use}
\label{tab:hallucination-mitigation-maturity}
\begin{tabular}{p{0.17\textwidth} p{0.44\textwidth} p{0.18\textwidth} p{0.16\textwidth}}
\hline
\textbf{Maturity Level} & \textbf{Observable Criteria } & \textbf{Severity Mapping} & \textbf{Representative P-IDs} \\
\hline
\textbf{Level 0: Ad-hoc} &
No shared policy; verification is up to individual intuition; analysts question it only if it feels off. &
None; reliance on individual vigilance. &
P02, P05, P14, P16 \\
\hline
\textbf{Level 1: Informal Norms} &
Shared human-in-the-loop culture; informal "never trust" agreements; manual cross-referencing is standard. &
Reactive human oversight. &
P01, P09, P10, P11, P12, P20  \\
\hline
\textbf{Level 2: Tactical Managed} &
Documented prompt libraries; task-specific templates; evidence-only prompting requirements. &
Documented preventive guardrails. &
P03, P07, P08 \\
\hline
\textbf{Level 3: Governed/Systemic} &
Automated validation scripts; schema validation; formal Legal/Compliance sign-off gates. &
Proactive, auditable technical controls. &
P04, P15, P17, P19  \\
\hline
\textbf{Not applicable } &
Organizational readiness could not be assessed due to lack of active LLM usage. &
N/A &
P06, P13, P18  \\
\hline
\end{tabular}
\end{table*}

\textcolor{black}{We include short de-identified exemplars illustrating the transcript evidence used to map maturity levels in Table \ref{tab:hallucination-mitigation-maturity}: Level 0 relies on individual intuition and occasional “spot-checking”; Level 1 reflects shared informal norms where LLM outputs are treated as leads and manually cross-checked; Level 2 uses documented, reusable prompt templates/libraries with evidence-only prompting; and Level 3 implements governed, auditable controls such as redundancy, structured outputs, and workflow-integrated validation steps.}





\begin{table*}[t]
\centering
\footnotesize
\color{black}
\setlength{\tabcolsep}{5pt}
\renewcommand{\arraystretch}{1.2}
\caption{Stratified analysis of LLM integration and operational risks across SOC roles.}
\label{tab:stratified-role-analysis}
\begin{tabular}{p{0.18\textwidth} p{0.20\textwidth} p{0.20\textwidth} p{0.18\textwidth} p{0.20\textwidth}}
\hline
\textbf{Role} & \textbf{Primary Tasks} & \textbf{Critical Failure Mode (Taxonomy)} & \textbf{Mitigation Priority} & \textbf{Design Implication} \\
\hline

Front-line Analysts (e.g., P01, P09, P10, P20) &
Log summarization, command-output synthesis, initial triage. &
FM1/FM3: Fabricated facts and misleading investigative direction. &
Human-in-the-loop: Treating output as a ``lead'' rather than a conclusion. &
Conciseness \& Citations: Focus on succinct summaries with direct links to raw log evidence to minimize verification drag. \\
\hline
Technical Engineers (e.g., P04, P05, P14, P17) &
Detection scripting, regex generation, logic translation. &
FM2: Invalid syntax/operators that break tool ingestion. &
Structured validation: Schema checks (JSON/YAML) and dual-control rule review. &
Syntax \& Schemas: Prioritize structured output formats and automated linting to prevent toolchain corruption. \\
\hline
Security Managers (e.g., P03, P11, P12, P19) &
SIEM strategy, unified dashboards, automated reporting. &
FM4: Interpretation errors leading to incorrect escalations. &
Governance gates: Legal/Privacy sign-offs and policy-driven oversight. &
Auditability: Implement verification logs and tiered data classification to manage compliance hurdles. \\
\hline
Tool Developers/Researchers (e.g., P07, P08) &
RAG implementations, prototype development, report synthesis. &
FM5/FM6: Opaque reasoning and speculative guessing under novelty. &
Architectural controls: Retrieval-Augmented Generation (RAG) and multi-model cross-checks. &
Transparency: Develop execution-trace dashboards and tiered explanation designs to demystify model rationale. \\
\hline
\end{tabular}
\end{table*}

\subsection{Taxonomy Coverage and Role Distribution}\label{c2}
\begin{table*}[htbp]
    \centering
    \footnotesize
    \caption{Taxonomy Coverage and Role Distribution ($N=20$)}
    \label{tab:taxonomy-coverage}
    \begin{tabular}{l c l p{0.3\columnwidth}}
        \toprule
        Failure Mode & Mentions & Primary Roles & Salience Summary \\
        \midrule
        FM1: Fabricated Facts & 11 & Analysts & Triage validity concern. \\
        FM2: Invalid Logic & 4 & Engineers & Toolchain disruption. \\
        FM3: Misleading Direction & 5 & Analysts & Forensic drag/wasted time. \\
        FM4: Incorrect Interp. & 6 & Analysts/Managers & Missed alert risk. \\
        FM5: Opaque Reasoning & 7 & Analysts/Managers & Verification bottleneck. \\
        FM6: Novelty/Sparse Intel & 4 & Researchers & Zero-day overconfidence. \\
        FM7: Verification Overhead & 4 & Analysts & Value-prop erosion. \\
        \bottomrule
    \end{tabular}
\end{table*}

\section{SOC Integration Constraint Matrix}\label{app:synthesis_matrix}
This appendix reports the decision-grade SOC Integration Constraint Matrix referenced in the main text. The matrix maps SOC task types to dominant failure modes (FM1--FM7), expected verification burden (VB), minimum maturity prerequisites (MP), and recommended safe integration patterns (SIP), using \textit{IE} where interview evidence is insufficient.
\begin{table*}[t]
\centering
\footnotesize
\color{black}
\caption{\textbf{SOC Integration Constraint Matrix.} Rows are SOC task types; columns map each task to dominant failure modes (FM1--FM7), expected verification burden (VB), minimum maturity prerequisite (MP; L0--L3), and a recommended safe integration pattern (SIP). Use \textit{IE} where transcript evidence is insufficient to justify a mapping. (FM definitions and L0--L3 rubric are provided elsewhere in the paper.)}
\label{tab:synthesis_matrix}
\setlength{\tabcolsep}{6pt}
\renewcommand{\arraystretch}{1.15}
\begin{tabular}{p{0.15\linewidth} p{0.26\linewidth} p{0.10\linewidth} p{0.10\linewidth} p{0.34\linewidth}}
\hline
\textbf{SOC task type} &
\textbf{Dominant failure modes} &
\textbf{VB} &
\textbf{MP} &
\textbf{Safe integration pattern (SIP)} \\
\hline

Summarization &
FM1, FM5, FM7 &
Med &
L1--L2 &
\textbf{Chat-only drafts:} treat output as a lead, not a conclusion; require evidence pointers (e.g., log lines, ticket IDs, or source references); include standard uncertainty disclaimers; keep human sign-off. \\

Detection drafting (rules/queries) &
FM2, FM5, FM7 &
Med &
L2--L3  &
\textbf{Template + schema validation:} structured output (YAML/JSON), linting/unit checks, and peer review; \textbf{Gated deployment:} promote to production only via CI + approval (dual-control). \\

Triage guidance (next steps / prioritization) &
FM3, FM4, FM1, FM6 &
High &
L2--L3 &
\textbf{Chat-only assist:} brainstorming with explicit uncertainty + alternatives; prohibit auto-execution; enforce “human final authority.” \textit{IE} for fully automated triage without human sign-off. \\

CTI synthesis (narrative intel, attribution context) &
FM1, FM5, FM6 &
High &
L3 &
\textbf{Gated deployment:} retrieval-grounded generation over trusted sources; require citations/evidence trail; cross-check workflow; audit logging for provenance and edits. \\

Script generation (automation/snippets) &
FM2 \(\pm\) FM5 &
Low &
L1--L2 &
\textbf{Chat-only drafts + execution tests:} run/verify in sandbox; lint + unit tests before reuse; optional templates for common tasks; never auto-run on production endpoints. \\

\hline
\end{tabular}
\end{table*}

\end{document}